# Cooperative *Phoneotypes*: Exploring Phone-based Behavioral Markers of Cooperation

Vivek K. Singh
School of Communication and Information,
Rutgers University, NJ, USA.

Rishav R. Agarwal,
Department of Economics,
Indian Institute of Technology,
Kanpur, India.

**ABSTRACT**
Cooperation is a fundamental human concept studied across multiple social and biological disciplines. Traditional methods for eliciting an individual's propensity to cooperate have included surveys and laboratory experiments and multiple such studies have connected an individual's cooperation level with her social behavior. We describe a novel approach to model an individual's cooperation level based on her *phoneotype* i.e. a composite of an individual's traits as observable via a mobile phone. This phone sensing-based method can potentially complement surveys, thus providing a cheaper, faster, automated method for generating insights into cooperation levels of users. Based on a 10-week field study involving 54 participants, we report that: (1) multiple phone-based signals were significantly associated with participant's cooperation attitudes; and (2) combining phone-based signals yielded a predictive model with AUCROC of 0.945 that performed significantly better than a comparable demography-based model at predicting individual cooperation propensities. The results pave the way for individuals and organizations to identify more cooperative peers in personal, social, and commerce related settings.

**Author Keywords:** Cooperation; Mobile Sensing; Behavioral Sensing; Phoneotype; Mobile Supported Cooperation

**ACM Classification Keywords:**
J.4. Computer Applications – Social and Behavioral Sciences.

**INTRODUCTION**
Cooperation allows humans to live together in societies and undertake tasks that cannot be undertaken by a single person. Thus, cooperation affects almost all aspects of human life including obtaining food, building houses, raising children, and doing commerce. As a result, cooperation has been very actively studied in multiple social and biological sciences (e.g. [43, 47, 54]). Such literature has identified multiple reasons for humans to cooperate (e.g. kin-based, reciprocity-based) and also studied the underlying processes as well as markers of individual propensities to cooperate in different settings (e.g. [29]).

This work focuses on identifying behavioral markers of cooperation. Specifically, we take inspiration from, and build upon a recent line of work on "cooperative *phenotypes*". The term "phenotype" refers to the observable physical properties of an organism which includes the organism's appearance, development, and behavior [44]. Recently, Peysakhovich et al. [43] have systematically studied and reported a long-term consistency in individual cooperation propensities that transcend the *state* that the individuals are in, and attributed this consistent propensity to cooperate to the existence of a "cooperative phenotype." While the breadth of observable characteristics of an individual remains large, in this work we focus on the set of characteristics observable via a mobile phone and call it a "*phoneotype*." Thus, we systematically explore the associations between phone-based behavioral markers and cooperation and quantify the predictive power of such data in inferring an individual's propensity to cooperate with others.

This approach differs diametrically from traditional methods for eliciting an individual's propensity to cooperate, which have mostly focused on traits that could be easily observed (e.g. gender, ethnicity) or elicited in a short time in laboratory settings. Unfortunately, the human-related information captured by observations in limited, unnatural settings must contend with multiple challenges, including subjectivity in observation, biases, and limited observation opportunities and at the same time involve substantive manual effort, cost and time [19]. Further, the reliance only on laboratory elicitable characteristics to identify cooperation propensities hampers the growth of the field of cooperation. Such an approach essentially rules out the potential for identifying behavioral markers based on communication or mobility traces spread over space and time to predict individual propensities to cooperate.

On the other hand, mobile phones and sensor-based data have been applied by researchers to create rich, personalized models of human behavior in social, spatial, and temporal contexts and connect them to outcomes like depression, happiness levels, and college GPA levels [8, 48, 53]. Thus, a phone-based approach toward predicting cooperation, if successful, could provide a cheaper, faster, scalable, and automated method for generating insights into cooperation levels of users.

Such a faster and automated method for inferring cooperation propensities has multiple practical implications. For example, much of the mobile and physical commerce (e.g. AirBnB or Craigslist) builds upon the expectation of cooperation from the other party in terms of putting in their fair share of effort/money in expectation of a mutually beneficial outcome. Ability to automatically infer the most cooperative peers would be very useful for individuals exchanging goods and services on such platforms. Further, given the chain reactions and contagion related to cooperation, it may be a useful tool for community designers to identify initial members in specific projects or cooperative endeavors (e.g. Wikipedia) so as to maximize the chances of success for the community [15, 55].

The contributions of this paper are two-fold.

*(1) To motivate and ground the use of phone-based signals for inferring cooperation propensities; and*

*(2) To define a machine learning model to automatically infer an individual's propensity to cooperate.*

The organization of the rest of the paper is as follows. We present the background and related work on the topic in the next section. This is followed by the description of a ten-week longitudinal study employed in this work. Next, we discuss the measures used in this study and explain the rationale and the details for the multi-faceted analysis undertaken. The obtained results and their implications are discussed before identifying the limitations and potential future work.

**BACKGROUND AND RELATED WORK**
Cooperation as a field of study has received intense interest across multiple disciplines (e.g. [54, 43, 4, 34, 2, 11]). In this section, we summarize the related work along four axes. First, we summarize the literature on cooperation as a field of study and pay specific attention to some of the key concepts (e.g. state vs. trait argument, and a summary of similarly used terms). Next, we discuss some of the literature connecting cooperation with sociability and mobility concepts (e.g. social capital, mobility capital, and strength of ties), that provides the theoretical motivation for this work. Next, we discuss some of the related work on markers of cooperation. Lastly, we summarize some of the related work in ubiquitous computing aimed at predicting different outcomes of interest regarding individuals based on the analysis of phone and other ubiquitous sensor data.

**Cooperation as a field of study**
Cooperation has received significant attention from the biological and evolutionary sciences [51, 15, 46]. For example, a biological explanation for cooperation may include the gain an individual receives through the impact on their own reproduction (direct) as well as the impact on related individuals (indirect) [51, 15]. Rilling et al. [46] provided a neural basis for cooperation stating that cooperation is associated with consistent activation in brain areas, which motivates humans to reciprocate favors.

There exist different schools of thoughts on the processes surrounding cooperation. Different explanations of its mechanics (e.g. as a trait vs. a state) exist [43, 4, 34], and the emerging literature suggests that neither of these can provide the complete picture, but rather cooperation is a complex composite of both the individual's stable traits and as well as the state they find themselves in [10]. A significant effort on understanding the traits facet of this equation was undertaken by Peysakhovich et al., in 2013 [43]. They posit that "Humans display a 'cooperative phenotype' that is domain general and temporally stable." While acknowledging the value of "states", the key finding of this work was that humans display a stable cooperative "trait", which just like "personality" transcends the immediate application scenario, and can explain human behavior across a wide variety of domains. Thus, the focus of this work remains on identifying the trait like mechanisms, which presumably manifest themselves in long-term patterns of human behavior as observed via mobile phones.

Another important distinction to make is between the related and often confused concepts of cooperation, collaboration, altruism, and pro-sociality. While altruism assumes a cost for the giver and benefit to the receiver; cooperation only predicates benefit to the receiver, the giver could also benefit from the transaction. We define these terms explicitly in our work as follows:

*Cooperation:* a behavior which provides a benefit to another individual (recipient), and which is selected for because of its beneficial effect on the recipient [54].

*Altruism:* a behavior which is costly to the actor and beneficial to the recipient; in this case, cost and benefit are defined on the basis of the lifetime direct fitness consequences of behavior [54].

*Collaboration:* Collaboration is a style of interaction between at least two co-equal parties voluntarily engaged in shared decision making as they work towards a common goal [21].

*Pro-Social Behavior:* Prosocial behavior can be defined as voluntary, intentional behavior that results in benefits for another; the motive is unspecified and may be positive, negative, or both [13].

While similar mobile-phone methods could be applied to study the other concepts in future, we maintain the focus of this work explicitly on predicting cooperation propensities of individuals.

**Cooperation and Socio-Mobile Behavior**
There exists a strong line of research connecting an individual's social behavior with her propensity to cooperate

[9, 35, 40]. Multiple theoretical and empirical efforts have also connected social and mobility related concepts with cooperation [45, 9, 40]. For example, a frequently cited concept connecting cooperation and social behavior is social capital. Social capital is defined as the "features of social organizations, such as networks, norms, and trust that facilitate action and cooperation for mutual benefit" [45]. In more empirical studies on the same thread, Curry and Dunbar [9] claimed that better-connected individuals are more likely to be cooperative; thus emphasizing the importance of social structure in explaining the patterns of cooperation and Molinas has used survey-based methods to find direct connections between social capital and cooperation levels [40].

The literature connecting mobility, i.e. a person's movements between locations throughout their daily activities, with cooperation has been relatively underdeveloped by contrast. While a user's position in the social network has been connected with cooperation and migration has been identified as a strategy to promote cooperation in spatial games [3, 9, 25], there are no formal theories connecting mobility and cooperation. On the other hand, foraging theory connects both mobility and cooperative hunting for food; human mobility has at least partly been attributed to social motivations [5], and the evidence connecting social capital with cooperation suggests that mobility capital [32] could also have some interesting associations with one's propensity to cooperate. While there is a lack of a formal theory connecting mobility and cooperation, we conjecture that this can at least in part be attributed to the lack of long-term mobility data being available to cooperation researchers in the past. Thus, some of the early links reported in this work may motivate more theoretical work in the near future.

**Markers of cooperation**
Researchers have also tried to explain cooperation by connecting it to various demographic features like age and gender [7] and laboratory elicitable psychological features like personality [33]. Biological and behavioral markers of cooperation have also been discussed in the literature [11, 47, 26]. However, while the approaches that focus on in-the-lab biological markers e.g. neurological and hormonal mechanisms for cooperation have often been mediated by technology (e.g. fMRI, hormonal measurements) [47, 11], the majority of the behavioral studies undertaken via ethnographic or survey-based study mechanisms and are yet to adopt automated methods to observe and predict human cooperative behavior [2, 26].

**Use of Mobile Phone to Predict Health and Wellbeing**
The emergence of mobile phones and sensors that people wear while undertaking activities of daily living is allowing researchers to create rich, personalized models of human behavior in social, spatial, and temporal contexts and connect them to outcomes like obesity, happiness levels, depression, financial outcomes, friendship, mental health, substance abuse, and college GPA levels (e.g. [8, 48, 53]). In fact, many of these studies have been grounded in past research connecting human behavior and life outcomes but have (re)gained prominence due to the availability of large-scale phone-based behavioral data to validate and refine the existing findings. Thus, leveraging this trend to predict cooperation levels, potentially opens doors for a newer method to study (and predict) the cooperation behavior of billions of users.

**STUDY**
This study probes the interconnections between cooperation levels and phone-based metrics based on a ten-week field + lab study involving 54 participants. The cooperation levels of the users have been obtained based on an in-lab survey and the phone-based metrics have been defined based on logs of call, SMS, and GPS data. The defined phone-based features have been analyzed in terms of their ability to predict the cooperation propensities using both regression and machine learning-based approaches.

**Participants**
The participant's for the study were recruited using flyers, email announcements, and social media posts in the area surrounding a major North American university. A total of 59 participants completed the study. Some of the participants did not complete all the surveys, and some did not enter their unique identifying code consistently across surveys, resulting in a set of 54 participants for whom we have the mobile-based data as well as the scores for the two surveys of interest (more details on surveys presented later).

Out of the 54 participants in the resulting dataset, 35 were male, and 19 were female. The most common age group was 18-21 years, most common marital status was single, the most common education level was "some college", and the median annual family income was in the range US $50,000-$74,999.

**Method**
The data used in this paper was obtained as part of the Rutgers Well-being study. In this study, the participants were invited to attend three in-person sessions which involved filling out several surveys pertaining to health, well-being, cooperation and some demographic data. The two surveys of relevant to the current paper are those on c*ooperation and demography.*

In the initial session, the participants were asked to read and sign a consent form to join the study and install a study client app onto their android cell phones. The study started with an orientation program where participants were trained on how to use the app.

Participants could opt out of the study at any point. The survey order was randomized for different participants. The participants were also instructed on the process of uninstalling the app and stopping data collection in the exit session.

All personnel involved in the study underwent human subject training and Institutional Review Board certification. In order to protect participants' personal information, all participants were identified using their phone's unique IMEI number which was kept constant throughout the course of the study. This unique identifier was anonymized via one-way cryptographic hashing before any analysis was undertaken.

The participants were compensated up to a sum of $100 ($20, $30, and $50 respectively for the three in-person sessions) on successful completion of the study.

**Data Collection**

Two types of information were collected from the participants for the ten week period.

*Communication log data (Call/SMS)*
This included date and time of communication event, the type of communication (text/call), incoming or outgoing, duration, and anonymized identifiers for both sender and receiver. Actual call audio and message text were **not** collected in this study.

*Physical location (GPS)*
This included the location of the device using GPS and approximate location area based on the cell tower. Such data were collected at an hourly resolution to trade-off between the battery consumption and the detail of coverage. This work focuses on GPS data that was analyzed as a <latitude, longitude> tuple at the fourth decimal point resolution. This roughly corresponds to 10m by 10m blocks and was selected as the scale since current GPS data have a pseudo-range accuracy of about 8 meters at 95% confidence level[1]. Since many of the participants spent inordinate amounts of times at the same location, while we were interested in characterizing their movement patterns, many of the features focused on the unique locations visited by the users.

**Survey Measures**

*Cooperation descriptor*
Prior literature on quantifying cooperation has identified multiple methods to observe the level of cooperation exhibited by an individual. These include games in strictly controlled lab setting e.g. dictator's game, prisoner's dilemma, etc. [2, 15, 35, 46], and surveys that elicit user behavior in certain described scenarios [43, 17]. While participant games are good at identifying user behavior (as opposed to *values*), they need to be tightly controlled, and there have been questions raised about their validity in predicting real-world (as opposed to lab-settings) behaviors [36]. Surveys, on the other hand, can explicitly differentiate between values and behaviors and be designed to ask users recent past behavior in practical settings (e.g. when was the last time you donated blood, gave money to a charity, gave money to a stranger, etc.) Such surveys though are critiqued to be theoretical and costly.

While leaving further resolution of such a debate outside the scope of this paper, we have decided to utilize a survey instrument for this work. As designing such a survey instrument to measure cooperation attitudes or behavior requires significant domain expertise (cooperation), we have adopted a measure of cooperation defined by Harvard scholars Peysakhovich, Nowak, and Rand. The survey described in a recent *Nature Communications* article has also laid the groundwork on the existence of a cooperation phenotype as already discussed earlier [43].

The survey consisted of 20 questions on a 5-point scale ranging from Strongly Disagree (1) to Strongly Agree (5), which were divided into two parts: 9 on cooperation values and 11 on cooperative behavior. An example of a cooperation value question is the level of agreement with "I would support an increase in taxes if it were used to help the less well-off in society", and an example of cooperation behavior question is to ask the user, "when was the last time they gave money to charity"? Note that the nature of these questions is not limited to a particular context and is explicitly focused on gathering an individual's propensity to cooperate in general everyday settings.

We focus on the total cooperation score that was defined as the sum of the scores obtained from the *values* and *behavior* questions. The full survey is available online[2]. While the theoretical maximum score obtainable for the survey was 100 (20 questions; 5-point scale), the observed maximum value was 80 and the median value was 59. A more detailed summary of the self-reported cooperation scores is shown in Table 1.

|  | **Minimum** | **Maximum** | **Median** | **Mean** |
|---|---|---|---|---|
| Cooperation Value | 27 | 43 | 37 | 36.83 |
| Cooperation Behavior | 14 | 40 | 22 | 23.3 |
| Total Cooperation | 44 | 80 | 59 | 60.13 |

**Table 1. Descriptive summary of self-reported cooperation survey data**

*Demographic Descriptors*
The participants were also asked questions about their demography (age, gender, marital status, the level of education, and family income level) using another survey.

**Smartphone data measures**
Based on a survey of existing literature we have identified a set of features to provide a representation of a user's social-mobile behavior. These features were in particular inspired by the literature on social capital [45] and its variants on "bridging" and "bonding" capital [56]. As suggested by Williams [56], we connect the concepts of "bonding" and "bridging" to those of "strong" and "weak" ties as proposed by Granovetter [21]. We have defined these features based on the working assumption that certain individual traits that manifest themselves in the long-term social and mobile behavior patterns of the users might be predictive of one's self-reported propensity to cooperate.

---

[1] http://www.gps.gov/systems/gps/performance/accuracy

[2] http://www.nature.com/ncomms/2014/140916/ncomms5939/extref/ncomms5939-s1.pdf

The features inspired by the "social" concepts have been operationalized using call and SMS data, and given the relative sparsity of the literature connecting mobility and cooperation, we consider physical movements to also be an extension of the social processes in the current discussion. After all, social processes and physical movements are intricately interconnected; meeting others is an important reason for one's movement patterns [5, 52], multiple research efforts have used GPS data to understand human and animal social processes [39], and human social process (e.g. friend-making) processes are moderated by the geographical constraints [41, 28].

**1. Social Activity Level (Call, SMS, GPS)**

$SA_i$ = Activity Count

Where activity is a new call/SMS exchanged, or GPS-location visited.

We measure the overall social activity by the number of calls, SMS, and unique locations experienced by a person over the course of the study. A high count would suggest that the person is more actively involved with people and places around her.

In our approach, we observe the physical location of each user once per hour. Hence counting the number of locations per day would yield exactly the same value (24) for every user. Hence, we focus on *unique* locations visited by the users. Given that meeting others is an important reason for one's movement patterns, we conjecture that a person visiting more unique locations is likely to have a more active social life [5, 52].

**2. Strong Ties Engagement Ratio (Call, SMS, GPS)**

$$S_i = \frac{\text{Communication count for the top third contacts}}{\text{Total communication}} \times 100$$

This quantifies the percentage of communication effort that a user devotes to her preferred contacts. This metric is based on two notions. (1) It utilizes the contact frequency as a proxy for tie strength as earlier adopted by Granovetter and others [20, 21, 31]. (2) It adopts a threshold of top-third as a cut-off for the "strong" ties. Similar top-third threshold has been adopted by multiple studies to demarcate the groups or the relationships with highest tie strength or social prestige in the past [22, 37]. In the current study, it is expected that a user will spend at least 33% of her time with her top third of contacts. However, a higher score (e.g. 80%) could be an indicator of a person's preference toward engaging significantly more with strong ties rather than spreading the interaction effort more equitably. We hypothesize that the relative spread (or concentration) of communication with such strong ties could be a predictor of one's propensity to cooperate.

Unique locations were considered as "contacts" for interpreting this feature for GPS data. We posit that a user visiting the same locations repeatedly does display a tendency to maintain strong connections with same preferred people or places.

**3. Weak Ties Engagement Ratio (Call, SMS, GPS)**

$$W_i = \frac{\text{Communication count for the bottom third contacts}}{\text{Total communication}} \times 100$$

This quantifies the percentage of communication effort that a user devotes to her less preferred (bottom third) contacts. A high weak tie engagement ratio would suggest that the user spends significant communication effort to interact with even the less preferred of his contacts. The contacts could be interpreted based on the call, SMS, or unique GPS locations.

**4. Diversity (Call, SMS, GPS)**

$$D_i = \sum_j p_{ij} \log_b p_{ij}$$

Where $p_{ij}$ is the percentage of engagements (Call/SMS/GPS) by individual 'i' to contact 'j', and 'b' is the total number of contacts.

The diversity score is based on Shannon Entropy and is a measure of how evenly a user's social engagements are distributed between different contacts. A user with low diversity distributes her communication unevenly across contacts, whereas a user with high diversity spends time evenly across many contacts. A similarly defined diversity metric has been found to be predictive of social mobility and economic well-being in different contexts in the past [12, 42]. Similarly, the diversity in locations visited has been reported as a fundamental characterization of animal foraging behavior [49] and found to be predictive of multiple life outcomes for animals including their social stature, survival rate, and reproductive success [23].

**5. Diurnal Activity Ratio (Call, SMS, GPS)**

Prior literature has connected animal rhythms and circadian cycles and cooperation [30]. The characterization of different individual's *chronotype* - the propensity for the individual to sleep at a particular time during a 24-hour period [27] – colloquially "morningness" or "eveningness" has been connected with cheating and Machiavellianism [27]. Lastly, the amount of sleep one obtains has been connected with one's cognitive abilities as well as cooperation levels [38, 1].

In this work we measure the circadian rhythms as the ratio of activity between two 12 hours "phases".

$$DiurnalRatio = \frac{\text{Communication count during phase 1}}{\text{Communication count during phase 2}}$$

We consider two natural definitions of "phases." First is the difference between "work" and "play". Based on discussions with some of the participants in the study, we decided to use 8 pm as the cut-off for this. Thus, we compare the activity between the period 8 am – 8 pm (a proxy for "work") with 8 pm – 8 am (a proxy for "play") and define a Diurnal8pmRatio.

$$DiurnalRatio8pm = \frac{\text{Communication count during 8am} - \text{8pm}}{\text{Communication count during 8pm} - \text{8am}}$$

Next, we wanted to capture the differences between mostly-awake and mostly-sleeping periods. Again based on an informal discussion with some of the participants (university students), who suggested 1 am as the most common time for students to go to sleep, we adopted 1 am as this threshold. Thus, we compare the activity levels between "day" i.e. mostly-awake period ending at 1 am, and "night" i.e. mostly-sleeping period to define a Diurnal1amRatio.

$$DiurnalRatio1am = \frac{\text{Communication count during 1pm} - \text{1am}}{\text{Communication count during 1am} - \text{1pm}}$$

The key difference between the two variants of the diurnal ratio is the period between 8 pm and 1 am, which was reported by the student participants to be an important period for social engagement and interaction in the student community. Thus a combination of the two diurnal scores might be a useful way to capture the circadian rhythms of the participants.

In both the interpretations, the individuals who have a diurnal activity ratio closer to 1 are less likely to prioritize communication or travel during specific hours of the day and spread out their social interactions unbiased of their personal circadian rhythms, while others give higher priority to their personal rhythms or have personal or social preferences resulting in different activity levels during different phases.

**6. In Out Ratio (Call, SMS)**

$$IOR = \frac{\text{Incoming communication count}}{\text{Outgoing communication count}}$$

It is defined as the ratio of the number of incoming Calls/SMS to outgoing Call/SMS messages. A higher In-Out ratio would suggest that the individual is more likely to receive Call/SMS and/or reply back to the Call/SMS received. This could connect with both the social capital and the reciprocity aspects of human cooperation. This feature has not been defined for GPS data.

A summary of the features defined is shown in Table 2 and a summary of the phone data collected in the study is shown in Table 3.

**Call/SMS/GPS**

| Metric | Formula | Description |
|---|---|---|
| Social Activity Level | Activity Count | Quantifies the overall social activity. |
| Strong Ties Engagement Ratio | $\frac{\text{Comm top} - \text{third contacts}}{\text{Total Comm}} \times 1$ | Quantifies the communication effort that a user devotes to her top third contacts |
| Weak Ties Engagement Ratio | $\frac{\text{Comm bottom third contacts}}{\text{Total Comm}} \times$ | Quantifies the communication effort that a user devotes to her bottom third contacts |
| Diversity | $\sum_j p_{ij} \log_b p_{ij}$<br>$p_{ij}$ = percentage of engagements made by individual 'i' to contact 'j', $b$ = total number of contacts. | Quantifies the evenness of social engagements. |
| Diurnal Activity Ratio | $\frac{\text{Comm during phase 1}}{\text{Comm during phase 2}}$ | Quantifies circadian rhythm – the ratio of communication taking place in two different 12hour phases. |
| In-Out Ratio | $\frac{\text{Incoming comm}}{\text{Outgoing comm}}$ | Quantifies the likelihood of replying to the incoming communication. |

Here ***comm*** is communication count.

**Table 2: Summary of metrics**

| Data type | Data points (54 participants) | Mean (each participant) | Median (each participant) |
|---|---|---|---|
| Calls | 27,988 | 518 | 314 |
| SMS messages | 186,677 | 3457 | 2486 |
| GPS (unique locations visited) | 14,338 | 266 | 281 |

**Table 3: Summary of mobile phone data considered in this study.**

The 54 participants considered in this study made a total of 27,988 calls and exchanged 186,677 messages during the 10 week period. This corresponds to an average of 52 (median = 31) calls per user per week, 346 (median = 249) text messages, and 27 (median = 28) unique locations visited per user per week. While neither of these captures the fullest details of an individual's interactions, the relatively high numbers indicate active usage of these phone-based features and suggest that long-term (rather than instantaneous) features derived from such data might be useful to capture the general trends in individual socio-mobile behavior i.e. *phoneotype*.

**RESULTS**

While certain applications may require prediction of exact numeric cooperation score, others might need to work with broader classifications e.g. "strong" or "weak" cooperators for the task at hand. Hence in order to support both these kind of applications, and to also obtain higher confidence in the observed results, we decided to undertake both these kinds of analysis.

**EXPLAINING THE VARIATION IN THE PROPENSITY TO COOPERATE**

We undertook linear regression analysis with the cooperation score as the output variable and the aforementioned demographic and socio-mobile variables as the input variables. Specifically, we tried three different models, one including only the demography variables, one including only the socio-mobile i.e. phoneotypic variables, and the third including the demography and socio-mobile variables. The demographic variables have been included as nominal variables resulting in dummy variables for each corresponding level. The method used for inclusion of variables was "Enter", which means that all variables were entered into the model at the same time. The implementation was undertaken using SPSS version 23.

**Table 4: Different regression models and their performance at explaining the variance in the propensity to cooperate. (n.s. = non-significant).**

|  | Variance Explained (Adjusted $R^2$) | Model Significance (p-value) |
|---|---|---|
| **Demography variables** | 0.017 | 0.341 (n.s.) |
| ***Phoneotype* (Socio-Mobile) variables** | 0.379** | 0.008 |
| **Demography + Socio-Mobile variables** | 0.498** | 0.006 |

As can be seen in Table 4 the model consisting of the demography variables (age, gender, marital status, the level of education, and family income level) yielded an adjusted R-square value of 0.017. We also undertook a corresponding ANOVA test to check the hypothesis that at least one of the corresponding coefficients in the regression equation was non-zero. The p-value obtained for the generated model was greater than 0.05 indicating that there was insufficient evidence to suggest that the coefficients for the demographics variables are non-zero. On the other hand, a model consisting of the socio-mobile i.e. phoneotypic features resulted in an adjusted R-square value of 0.379 (p-value =0.008) indicating that one or more of the coefficients are likely to be non-zero. Lastly, a model consisting of both socio-mobile and demography variables resulted in an adjusted R-square value of 0.498 (p-value =0.006).

This means that 49.8 % of the variance in the participant's cooperation score was explained by demography and phone-based signals. While there is no universal metric for judging the quality of this score, we note that 0.498 would fall between "moderate" and "substantial" based on interpretation suggestions in the existing literature [14, 6]. To further interpret this result, we note that percentages of variation in individual height and body mass explained by one's genes are 55% and 23% respectively [57].

The low score for the "only demography" model indicates that the demographic features by themselves did not explain significant variance in the cooperation levels. However, they were useful in increasing the adjusted R-square score for the phoneotypic model. This increase also suggests that phoneotypic features and demography features are not merely proxies for each other but rather add newer information.

**PREDICTIVE CLASSIFICATION MODEL FOR LEVEL OF COOPERATION**

To build and test a classification approach for cooperation propensities, we divided the cooperation scores based on the median value (59) of the sample. The median value of the sample was chosen as there are no "standardized" median scores available for the selected survey instrument. This resulted in two classes, which we refer to as "Strong" cooperators (N=26) and "Weak" cooperators (N=28) in our discussion here.

To build the predictive models we first undertook feature selection based on optimal subset selection method as proposed in [50]. This method evaluates the worth of a subset of attributes by considering the individual predictive ability of each feature along with the degree of redundancy between them. Subsets of features that are highly correlated with the class while having low inter-correlation are preferred. We used the implementation of this method as included in Weka version 3.6 [24].

**Table 5: Features selected for the different models considered.**

| Approach | Features Selected |
|---|---|
| **Only Demography** | Age, Marital Status |
| **Only *Phoneotype*** | Diurnal1am.location.ratio, Diurnal8pm.location.ratio, Diurnal1am.Call.ratio |
| **Demography + *Phoneotype*** | Diurnal1am.location.ratio, Diurnal8pm.location.ratio, Diurnal1am.Call.ratio, Marital Status |

| Method | Only Demography | | Only Phoneotype (Socio-Mobile) | | Phoneotype (Socio-Mobile) + Demo | |
|---|---|---|---|---|---|---|
| | AUCROC | Accuracy | AUCROC | Accuracy | AUCROC | Accuracy |
| **AdaBoost** | 0.5 | **66.7** | **0.945** | 74.1 | **0.945** | 74.1 |
| **LogitBoost** | **0.502** | **66.7** | 0.794 | 75.9 | 0.782 | 74.1 |
| **BayesNet** | 0.5 | 64.8 | 0.819 | **79.6** | 0.819 | **79.6** |
| **NaiveBayes** | 0.5 | 64.8 | 0.753 | 64.8 | 0.753 | 64.8 |
| **RandomTree** | 0.5 | 64.8 | 0.629 | 63 | 0.742 | 74.1 |
| **Zero R** | 0.5 | 51.8 | 0.5 | 51.8 | 0.5 | 51.8 |

**Table 6: Prediction accuracy for cooperation levels using different approaches**

Similar to the regression analysis, we considered three different alternative models based on only demographic variables, only socio-mobile (phoneotypic) variables, and demographic + phoneotypic variables. The process of features selection yielded a much smaller set of features for the three models as shown in Table 5.

These selected features were passed to different machine learning algorithms for classification. Specifically, we used five well-established methods Adaptive Boosting (AdaBoost), Logistic Boost (LogitBoost), Bayes Network, Naïve Bayes, and Random Tree as the classification methods and decided to adopt a leave-one-out cross validation method to a tradeoff between the learning opportunities and the generalizability of results from the data. Table 6 presents a summary of the results obtained in terms of accuracy as well as the AUCROC (area under the Receiver Operating Characteristic curve) scores. Note that AUCROC can also be a useful metric in classification scenarios when a trade-off between true positive rate and false positive rate is of vital interest. (Note that the baseline for ROC was taken to be 0.500 irrespective of the cross-validation.)

As shown in Table 5, the demography-based model yielded the best accuracy of 66.7% and best ROC of 0.502. The only *phoneotype* based model yielded a best case accuracy of 79.6% and AUCROC of 0.945. While there were minor differences between the performances of different classification algorithms between only *phoneotype* and *phoneotype* + demography based models, the best case accuracy (79.6%) and AUCROC (0.945) remained the same.

To aid interpretation of the results, we also report the results based on a baseline 'Zero-R' approach, which simply classifies all data into the largest category. Given the nearly equal distribution of classes, the baseline Zero-R approach resulted in 51.8% accuracy and 0.500 ROC in the classification tasks (without cross-validation). Phone based socio-mobile features could correctly classify the cooperation level 79.6% (resp. 0.945 AUCROC) of the times, which marks a 58.2% (resp. 89% in terms of AUCROC) relative improvement over the baseline method and a 19.3% (respectively 88.2% in terms of AUCROC) improvement over the demography based model.

This corroborates the findings from the regression analysis, that phone-based features significantly outperform demography-based features in prediction accuracy, and perform well both with and without including the demography based features. A possible interpretation for this, is that going beyond the traditional demographic descriptors, the availability of rich mobile sensing data (e.g. calls made, SMS sent, and unique location visited over 10 week period), allowed for creation of a more detailed model for individual behavior, yielding a higher predictive power on cooperation propensities.

Another interesting point to note is regarding the features selected. First, it was interesting to note that all three features selected from the *phoneotype* data were based diurnal activity ratios. Such an interconnection has been alluded to in different biological and psycho-social contexts [30, 1], but it has been directly connected with phone-based movement and calling patterns for the first time in this work.

This suggests that future research leveraging ubiquitous sensing data might have a useful role in expanding the understanding the interconnections between socio-mobile behavioral data and cooperation propensities.

To gain further insight into the phoneotypic features identified and their relative effect on the propensity to cooperate we undertook posthoc correlation analysis between the cooperation score and the different phoneotypic features identified. A summary of the partial Pearson's correlation coefficients after controlling for demographic variables is shown in Table 7.

As can be seen in Table 7, the most significant predictor for cooperation propensity was found to be Diurnal Ratio 8pm (GPS) (r=-0.447; p-value= 0.002). As Diurnal Ratio is the ratio of unique locations visited during "work" to the unique locations visited during "play", a possible interpretation of this result is that individuals who have a heavier concentration of traveling during the "work" period compared to the "play" period are less likely to be cooperative.

The next highest scoring feature was the number of phone calls made by the person, Social Activity Level (call), (r=.388; p-value= 0.009). This corroborates well with existing literature that suggests a link between social behavior and propensity to cooperate [9, 45]. This work is the first to use the number of phone calls (presumably a proxy for social capital) to quantitatively demonstrate this connection, though.

**Table 7: Pearson's correlation coefficients for the different *phoneotypic* features considered after controlling for demography features. n.s. = not significant; Significance codes are as follows: 0 '**' 0.01 '*' 0.05 'º' 0.1 '' 1. Green indicates a positive correlation, Red negative.**

| Phoneotypic Features | Partial correlation coefficient | Significance |
|---|---|---|
| Social Activity Level (Call) | 0.388** | p=0.009 |
| Social Activity Level (SMS) | 0.091 | n.s. |
| Social Activity Level (GPS) | -0.165 | n.s. |
| Strong Ties Engagement (Call) | 0.126 | n.s. |
| Strong Ties Engagement (SMS) | 0.274º | p=0.072 |
| Strong Ties Engagement (GPS) | -0.106 | n.s. |
| Weak Ties Engagement (Call) | -0.108 | n.s. |
| Weak Ties Engagement (SMS) | -0.100 | n.s. |
| Weak Ties Engagement (GPS) | 0.082 | n.s. |
| Diversity (Call) | 0.122 | n.s. |
| Diversity (SMS) | 0.044 | n.s. |
| Diversity (GPS) | 0.120 | n.s. |
| Diurnal Ratio 1 am (GPS) | 0.044 | n.s. |
| Diurnal Ratio 8pm (GPS) | -0.447** | p=0.002 |
| Diurnal Ratio 1 am (Call) | 0.304* | p=0.045 |
| Diurnal Ratio 8pm (Call) | -0.136 | n.s. |
| Diurnal Ratio 1 am (SMS) | 0.198 | n.s. |
| Diurnal Ratio 8pm (SMS) | -0.132 | n.s. |
| In Out Ratio (Call) | 0.030 | n.s. |
| In Out Ratio(SMS) | -0.158 | n.s. |

The third highest scoring feature is Diurnal Ratio 1am (call) (r=0.304; p-value= 0.045). As this feature is the ratio of the number of phone calls made during mostly-awake or "day" phase to the number of phone calls made during "night" phase, this suggests that individuals who have a heavier concentration of phone calls during the "day" period compared to the "night" period are be more likely to be cooperative. This can be loosely connected to the chronotype literature suggesting that individuals who are more active during the night are less likely to cooperate [1]. Further analysis and interpretation are left as part of future work.

The only other phoneotypic feature with (marginally) significant correlation was Strong Ties Engagement Ratio (SMS) (r=0.274; p-value= 0.072). The positive association between the relative engagements with "strong ties" is also along expected lines based on the literature connecting social capital (bonding capital) and cooperation levels [56, 40]. We also note that while the underlying mechanics for subset feature selection in classification and Pearson's correlation analysis are quite different, two of the four features found significant via correlation analysis were also part of the subset selected via the "optimal subset selection" method used for creating classification prediction models.

## DISCUSSION

### Methodological Considerations

Here we discuss multiple design choices made and the corresponding caveats concerning the three types of analysis (correlation, regression, and classification) undertaken in this work.

First, we note the multiple comparisons undertaken in the correlation analysis. While such multiple comparisons are often "corrected" using Bonferroni or Bonferroni-Holm correction to maintain the confidence in the associations found, we do not do so in this work. This is because the analysis undertaken here is posthoc and intended to help interpret the observed prediction results rather than being prescriptive in its own right. Similarly, we acknowledge the issues associated with the use of a relatively large number (21) of possibly collinear features in regression given the modest sample size. While this makes the interpretation of individual features difficult, the overall explanation scores of 0.379 for *phoneotype* (respectively 0.498 for *phoneotype + demography*) remain interpretable.

Given the limited sample size, we focus on triangulating and identifying general trends across the three analysis methods (correlation, regression, and classification) rather than establishing hard associations between specific variables. The common question tying the three threads of analysis is: *can socio-mobile signals as observed via a phone be used to infer the cooperation propensity of an individual?* The answer to this question based on the general consistency of results across the threads of analysis is *positive*.

### Privacy of User Data and Ethical Considerations

All data used in this study were hashed and anonymized as discussed in the study design. At no point were the actual phone numbers or the content of the call or SMS text available to the personnel undertaking analysis.

The permissions needed for the study (call logs, SMS logs, location logs, and phone identifier information) were designed to be significantly lesser than those typically adopted by popular apps. For example, the Facebook app for Android requires permission to: read phone status and identity, directly make phone calls, read SMS/MMS messages, take pictures and video, record audio, precise and approximate location, read and modify contacts, read and edit the calendar, etc. Lastly, the participation in the study was on a voluntary basis, and the participants could drop out at any time in the 10-week study.

We also note the ethical concerns surrounding assigning an individual a score based on her propensity to cooperate. While such scores could be used by an individual to demonstrate their suitability to potential partners in different transactions (e.g. CouchSurfing, Swap.com), sometimes the benefits may accrue to the society rather than the individual. Similar concerns have been raised about the traditional paper survey based methods with a similar goal, and also newer automated techniques that use social media and phone data to assign health, wellbeing, or similar "suitability" scores to individuals [34]. Instead of shunning away from reporting such results, we adopt the approach of raising awareness about these new possibilities and informing the policy debate surrounding them.

### Limitations

The current study also has some limitations. First is the homogeneity of the sample. While this limitation prevents us from generalizing the findings to larger populations, the homogeneity also allowed us to isolate socio-mobile behavior as a predictor. Here we are able to focus on one group and predict relative variation in their cooperation levels as a function of their socio-mobile behavior. A second limitation is the relatively small sample size - 54 individuals. We also note that none of the results obtained here are causal, and hence the directionalities of the explanation could indeed be inverse. Taking into account these limitations, we will be cautious in generalizing the results obtained until they are verified at scale over representative sample populations.

Despite these limitations, this study is the first of its kind. To our knowledge, there have been no previous studies undertaken that analyze the link between cooperation level and phone-based socio-mobile behavior (*phoneotype*). The obtained results are encouraging, and have demonstrated the ability of phone-based social signals at predicting cooperation levels of individuals.

### Implications

The results open the doors to a methodology that, with refinements and validation, could be used at scale. Mobile phones are now actively used by more than 3 billion users, and hence the proposed method could potentially be applied to estimate the cooperation levels for billions of individuals.

In future, this work could also have multiple implications for social scientists, economists, mobile phone service providers, and policy designers. For example, these could be used by application designers (in say a crowd-sourcing task) for better participant acquisition (e.g. identifying participants more likely to cooperate with others), engagement and community development. Individual users may also benefit from obtaining and sharing such cooperation propensity scores as an alternate set of credentials in social tasks involving collective intelligence, and various facets of mobile commerce and the shared economy e.g. Uber, CouchSurfing and Task Rabbit.

### CONCLUSIONS AND FUTURE WORK

As an alternative to survey-based methods, which are often costly, labor-intensive, and time-consuming, this work motivates and grounds the use of phoneotypic (i.e. phone based socio-mobile) descriptors for inferring the cooperation propensities of different individuals.

Building upon social capital and foraging theory literature, this work has identified a number of socio-mobile features that may have predictive power on cooperation and analyzed the relevant associations. The identified phone-based features were combined into a predictive model that achieved AUCROC performance of 94.5% at two-way classification and performed significantly better than a comparable demographic model. The obtained results open significant opportunities for both studying and predicting cooperation levels at the scale of billions of users. For example, such results could be used by an individual to identify the most cooperative peers for engaging in mobile commerce or by community designers to recruit suitable early volunteers.

We plan to expand the scope of this study into a "living lab" setting in the near future, where we can engage with a much larger representative population to study similar questions based on their cooperation behavior in day to day activities.

### ACKNOWLEDGEMENTS

We would like to thank Cecilia Gal, Padmapriya Subramanian, Ariana Blake, Suril Dalal, Sneha Dasari, and Christin Jose, for help with conducting the study and processing the data. We would also like to thank Jacopo Staiano (LIP6), and Neal Lathia (Skyscanner) for very helpful feedback on an earlier draft of this paper.